\documentclass[pra,10pt,twocolumn,groupedaddress,floatfix,showpacs]{revtex4-1}

\usepackage[scaled=0.86]{berasans}  
\usepackage[colorlinks=true, allcolors=blue, urlcolor=blue]{hyperref}  
\usepackage{graphicx} 
\usepackage[babel]{microtype}  
\usepackage{amsmath,amssymb,amsthm,bm,amsfonts,mathrsfs,bbm} 

\usepackage{xspace}  
\usepackage{pgfplots}
\usepackage{xcolor,colortbl}
\usepackage{array}
\usepackage{bigstrut}
\usepackage{tcolorbox}
\usepackage{algorithm} 
\usepackage{algpseudocode} 
\algnewcommand{\To}{\textbf{to }}
\algnewcommand\Input{\item[\textbf{input:}]}%
\algnewcommand\Output{\item[\textbf{output:}]}%

\algdef{SE}[DOWHILE]{Do}{doWhile}{\algorithmicdo}[1]{\algorithmicwhile\ #1}%

\newcommand{\be}{\begin{equation}}
\newcommand{\ee}{\end{equation}}
\newcommand{\bea}{\begin{eqnarray}}
\newcommand{\eea}{\end{eqnarray}}
\newcommand{\bes}{\begin{equation*}}
\newcommand{\ees}{\end{equation*}}
\newcommand{\beas}{\begin{eqnarray*}}
	\newcommand{\eeas}{\end{eqnarray*}}

\newtheorem*{thm*}{Theorem}

\newtheorem*{lem*}{Lemma}

\newtheorem*{lipschitzLem*}{Lemma \ref{lipschitz}}
\newtheorem*{lipschitzCubeLem*}{Lemma \ref{lipschitzCube}}
\newtheorem*{pgmNearlyOptimalThm*}{Theorem \ref{pgmNearlyOptimal}}

\begin{document}


\title{  Quantum Amplitude Amplification Operators }


\author{Hyeokjea Kwon }
\author{Joonwoo Bae}
\affiliation{School of Electrical Engineering, Korea Advanced Institute of Science and Technology (KAIST), 291 Daehak-ro, Yuseong-gu, Daejeon 34141, Republic of Korea.}


\begin{abstract}


In this work, we show the characterization of quantum iterations that would generally construct quantum amplitude amplification algorithms with a quadratic speedup, namely, quantum amplitude amplification operators (QAAOs). Exact quantum search algorithms that find a target with certainty and with a quadratic speedup can be composed of sequential applications of QAAO: existing quantum amplitude amplification algorithms thus turn out to be sequences of QAAOs. We show that an optimal and exact quantum amplitude amplification algorithm corresponds to the Grover algorithm together with a single iteration of QAAO. We then realize $3$-qubit QAAOs with the current quantum technologies via cloud-based quantum computing services, IBMQ and IonQ. Finally, our results find that fixed-point quantum search algorithms known so far are not a sequence of QAAOs, e.g. the amplitude of a target state may decrease during quantum iterations.
\end{abstract}

\maketitle

\section{Introduction}

Recent advances with the noisy intermediate-scale quantum (NISQ) technologies \cite{Preskill2018} signify the usefulness of iterations of a quantum circuit that may contain advantages over its classical counterpart \cite{Bravyi308}. Then, the iterations may be used to devise hybrid quantum-classical algorithms toward more efficient computation over the existing classical limitation \cite{farhi2014quantum, Peruzzo:2014uz}. For instance, it is of great importance to find the capabilities of parametric quantum circuits exploited in variational quantum algorithms, see e.g., \cite{Benedetti_2019, PhysRevResearch.2.033125, https://doi.org/10.1002/qute.201900070}. 

This structure is in fact shared in common with Grover's quantum database search algorithm or quantum amplitude amplification in general \cite{PhysRevA.98.062333}. Namely, a Grover iteration is a building block that is repeatedly applied and leads to a quadratic speedup over classical search \cite{PhysRevLett.79.325, Bennett:1997wy}. We also recall that the optimality proof of the algorithm is concerned with the capability of single iterations whereby the speedup is achieved by their cancatenation \cite{PhysRevA.60.2746, Boyer:1998wm}.

There are two shortcomings in the Grover iterations, as follows. The first is that there exists an error in a measurement readout stage though being sufficiently small up to $1/2^n$ for $n$-qubit states. This simply means that the Grover algorithm finds a target by allowing  a non-zero error. It has been shown that the errors can be cleared out by improving the angle parameters in Grover iterations: then, exact quantum search that finds a target with certainty is achieved \cite{PhysRevA.64.022307, PhysRevA.62.052304, Toyama:2013te}. The second is that when a target is multiple, the number should be provided in advance. Otherwise, the algorithm shows oscillations between an initial and a target states \cite{PhysRevLett.79.325, Bennett:1997wy}. Or, quantum counting is required as a prescription \cite{MR1947332}, which can also be simplified \cite{Aaronson:2019wy, Suzuki:2020ts}. 

Then, fixed-point quantum search has been proposed in Ref. \cite{PhysRevLett.95.150501}, showing that a given initial state can arbitrarily converge to a target one. An {\it a priori} information about the number of target items is not necessarily known in advance. A drawback is that the quantum advantage with a quadratic speedup is not attained. Then, it has been shown that a fixed-point search algorithm can be obtained with a quadratic speedup \cite{PhysRevLett.113.210501}. However, as soon as exact search is attempted, it is inevitable that a quadratic speedup is ruled out \cite{PhysRevLett.113.210501}. One can summarize that fixed-point quantum search also contains a non-zero error probability in a measurement readout. 

Therefore, the shortcomings are not overcome simultaneously yet so far. Exact and fixed-point quantum search is not yet achieved. It may be observed that quantum amplitude amplification is generally composed by quantum iterations. We here ask the characterization of quantum iterations such that their sequences may lead to a quadratic speedup in general. The identification may also give the close-up view to find how exact and fixed-point quantum search are distinct. From a practical point of view, the iterations would be also a useful building block in the NISQ algorithms, as quantum amplitude amplification is a key process in various quantum computing applications, e.g., optimization \cite{10.1137/040605072, optimizaiton}, state preparation \cite{Soklakov:2005wx}, high-energy physics \cite{PhysRevD.101.094015}, cryptanalysis \cite{8490203, 10.1007/978-3-319-29360-8_3, PMCID:PMC7991734}, etc. Moreover, the process turns out be a naturally occurring phenomenon \cite{PhysRevLett.124.180501}. 


\begin{figure*}[t]
	\begin{center}
		\includegraphics[angle=0, width=.7 \textwidth]{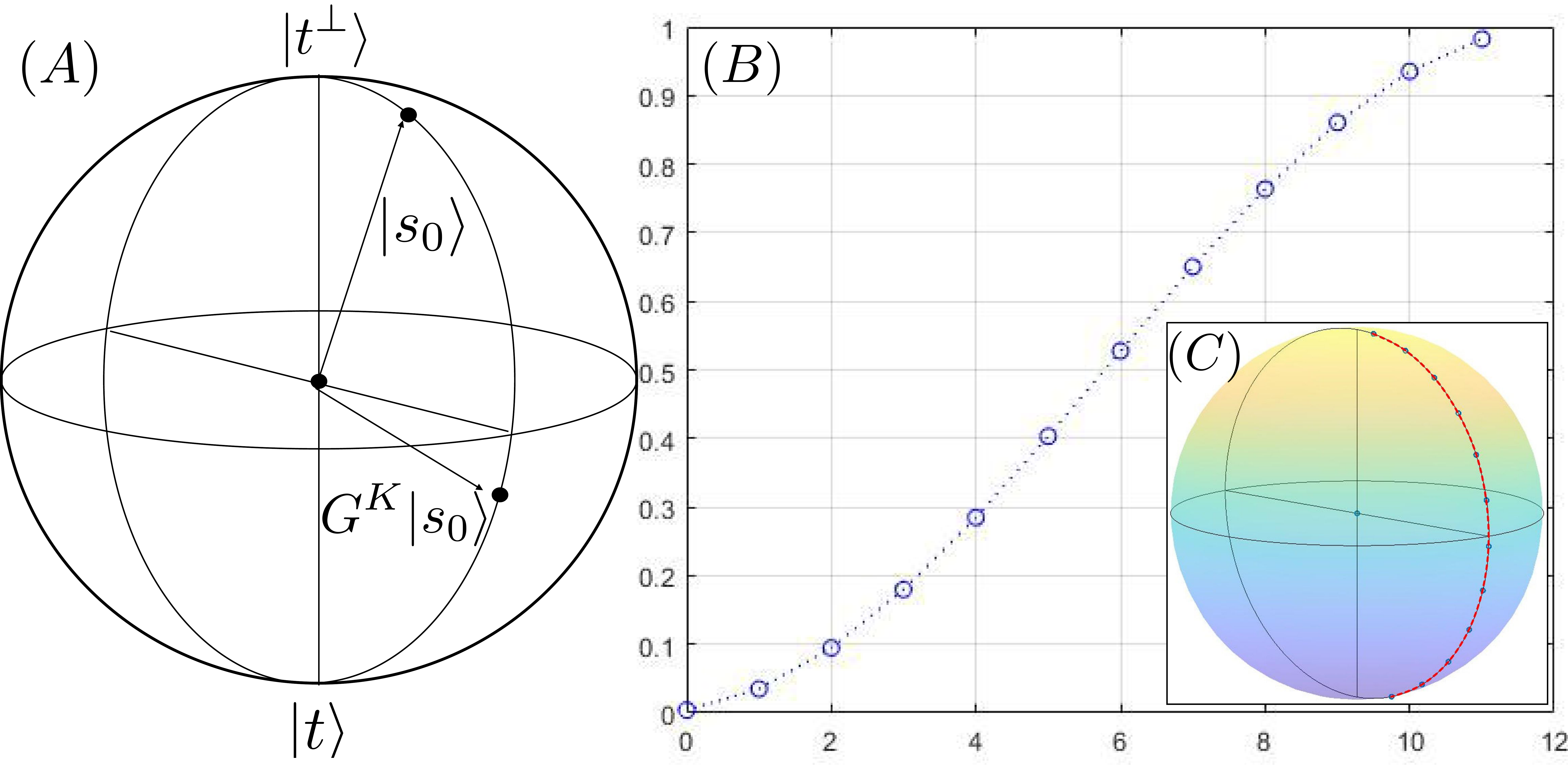}
		\caption{ (A) The Grover iteration corresponds to consecutive rotations in the space spanned by a target state $|t\rangle$ and its complement $|t^{\perp} \rangle$. (B) The probability of finding a target state is plotted in the case of $8$ qubits. The probability is monotonically increasing. (C) The path of an evolving state in the sphere is shown by Grover iterations from an initial to target states. } \label{grover}
	\end{center}	
\end{figure*}

Along the line, one can see that the Grover iteration has two subroutines, an oracle query and a specified diffusion operations. There is a high chance that a diffusion step in the Grover iteration may be served by any unitary transform \cite{PhysRevLett.80.4329}. This readily means that quantum amplitude amplification may be characterized in a wide range of parameters while the oracle query operation is performed in a noise-free manner. 
 
In this work, we show the characterization of quantum iterations that can be generally used to construct quantum amplitude amplification with a quadratic speedup. Namely, quantum amplitude amplification operators (QAAOs) are identified. It is shown that QAAOs can be obtained in a wide range of parameters: randomly generated parameters can build a QAAO with a probability almost $1/2$. This means that in practice QAAOs can be straightforwardly generated, and also that QAAOs are generically resilient to errors appearing in the preparation of the parameters. We realize QAAOs in the cloud-based quantum computing services of IBMQ and IonQ, and show that a single iteration of a QAAO can be realized with NISQ technologies.


\section{Quantum Amplitude Amplification Operators}

In this section, we present the characterization of QAAOs for $n$-qubit states. For convenience, we consider amplification of the amplitude of a single target state. Let $|s_0\rangle$ denote an initial state, 
$|t\rangle$ the target one, and $|t^{\perp}\rangle$ the orthogonal complement to the target state. The initial state is often given as a uniform superposition of $N=2^n$ states, 
\bea
|s_0\rangle = \frac{1}{\sqrt{N}} \sum_{j=0}^{N-1} |j\rangle = \frac{1}{\sqrt{N}}|t\rangle + \sqrt{\frac{N-1}{N}} |t^{\perp}\rangle.  \nonumber
\eea
In general, an $n$-qubit state in the space spanned by $\{|t\rangle, |t^{\perp}\rangle \}$ can be written as
\bea
|s(\theta,\phi) \rangle = e^{i\phi}\sin  ( \frac{\theta}{2})  |t\rangle   + \cos  ( \frac{ \theta}{2})  |t^{\perp}\rangle  \label{eq:state}
\eea
for some $\theta$ and $\phi$, see Fig, \ref{grover}. For a state in Eq. (\ref{eq:state}) the probability of finding a target is given by
\bea
p (\mathrm{target}) = | \langle t | s(\theta,\phi)\rangle |^2 = \sin^2 \frac{\theta}{2}. \label{eq:tprob}
\eea
The initial state can also be written as,
\bea
|s_0\rangle = |s (\theta_0, \phi_0=0)\rangle~\mathrm{with}~\theta_0 = 2 \sin^{-1}\frac{1}{\sqrt{N}}. \label{eq:initial}
\eea
Note that the initial state can be prepared by applying Hadamard gates to $n$ qubits prepared in a state $|0\rangle^{\otimes n}$.

\subsection{ Quantum iteration}

\begin{figure*}[t]
	\begin{center}
		\includegraphics[angle=0, width=0.7 \textwidth]{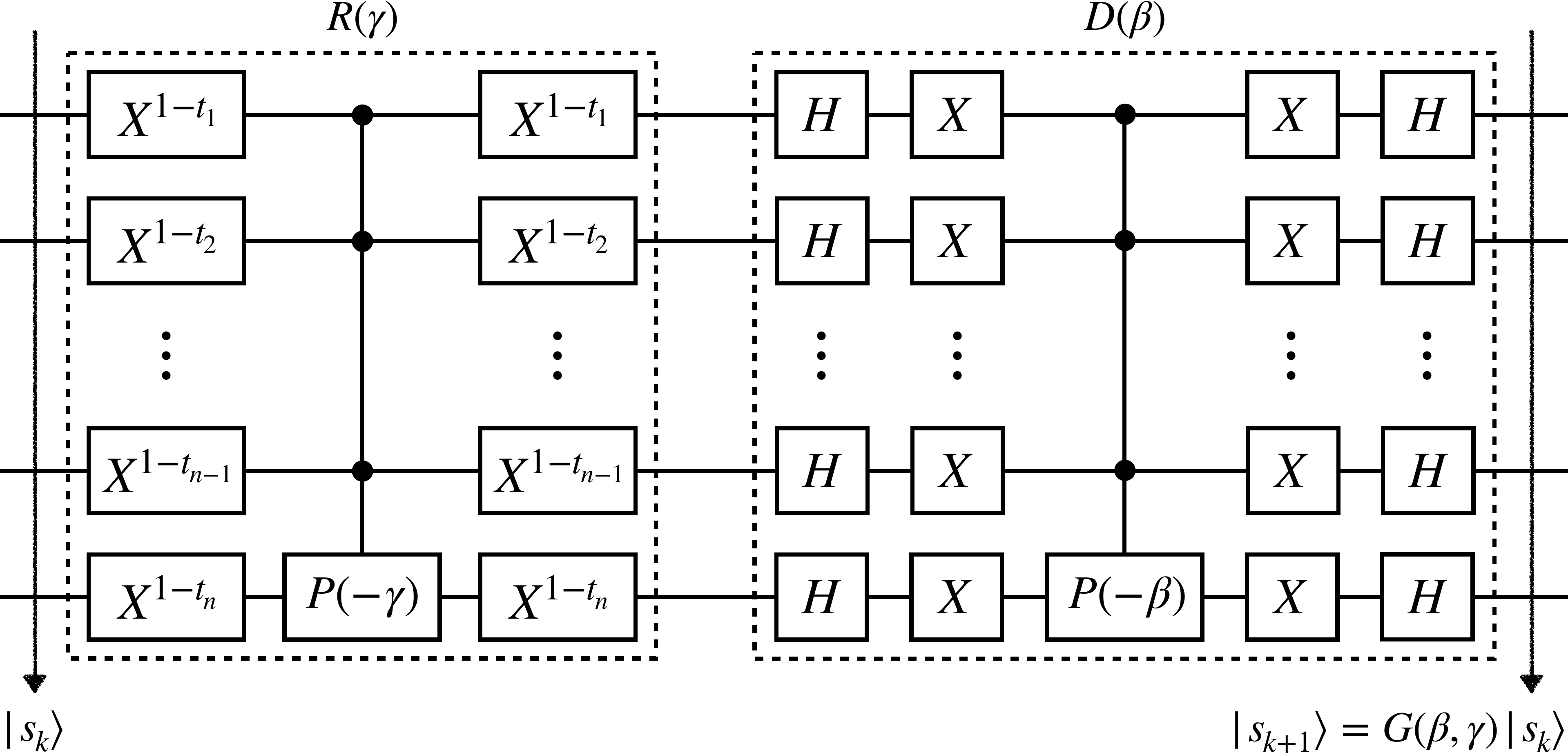}
		\caption{ A circuit of an iteration by a QAAO is shown, where a target and an initial states are denoted by $|t\rangle =|t_1 t_2 \cdots t_n\rangle$ and $|s_0\rangle$, respectively. An oracle gate $R(\gamma)$ applies a controlled-phase gate that realizes the transformation, $|t\rangle\mapsto e^{-i\gamma}|t\rangle$, for a target state only. A diffusion gate $D(\beta)$ is constructed with the {\it a priori} information about an initial state. Note that a phase gate is denoted by $P(\alpha) =\mathrm{diag}[1,e^{i\alpha}] $.
		   } \label{circuit}
	\end{center}	
\end{figure*}


Let us begin by identifying the parameters to construct a quantum iteration that leads to a quadratic speedup in amplitude amplification. It is not difficult to see that the quantum iteration corresponds to a rotation in the space spanned by a target state and its orthogonal complement $|t \rangle$ and $|t^{\perp}\rangle$, respectively. It then follows that a sequence of quantum iterations realizes a transformation toward a target state, and leads to a sufficiently high probability to find a target state. 



Thus, a quantum iteration can be realized in a decomposition as follows,
\bea
&& G(\beta,\gamma) = D(\beta) R(\gamma),~\mathrm{where}~ \nonumber \\
&& D(\beta) =e^{-i\beta|s_{0}\rangle\langle s_{0}|}~\mathrm{and} ~R(\gamma) =e^{-i\gamma|t\rangle\langle t|} \label{eq:io}
\eea
for $\beta,\gamma \in [-\pi,\pi]$. Note that the operation $D(\beta)$ called a diffusion can be constructed with an initial state given from the beginning, see Eq. (\ref{eq:initial}). The other one $R(\gamma)$ is called an oracular operation based on an oracle query, which is a one-way function $f(x) = \delta_{t,x}$ for $x \in \{ 0,1\}^n$. Then, the oracle operation works as, $U|x\rangle|y\rangle = |x\rangle |y\oplus f(x)\rangle$ for $x,y \in \{ 0,1\}^n$. In Fig. \ref{circuit}, a circuit for the operation $R(\gamma)$ is shown. Note that the Grover iteration corresponds to the iteration with $(\beta,\gamma) =(\pi, \pi)$. Note also that when $\gamma=\pi$, almost any unitary transformation may serve a diffusion step \cite{PhysRevLett.80.4329}.

Let $|s_j\rangle := |s (\theta_j, \phi_j)\rangle$ denote a state obtained after $j$ iterations, for which the probability of finding a target is given by 
\bea
p_j: = p_j (\mathrm{target}) =  \lvert \langle t |s_j \rangle \rvert^{2}= \sin^{2}\frac{\theta_j}{ 2}. \nonumber
\eea  
An increment by the next iteration $G(\beta_{j+1},\gamma_{j+1})$ for a given state $|s_j\rangle$ is given as follows,
\bea
\triangle_{p_j} (\beta_{j+1}, \gamma_{j+1})  
&:=& p_{j+1} -p_j \label{eq:probj} \\
&=& \lvert \langle t| G(\beta_{j+1}  ,\gamma_{j+1}  ) |s_j  \rangle \rvert^{2} -  \lvert \langle t |s_j \rangle \rvert^{2}.~~ ~~\label{eq:inc}
\eea
Note that the increment depends on parameters $(\beta_{j+1}, \gamma_{j+1})$ of the $(j+1)$-th iteration and the target probability $p_j$ of a given state $|s_j\rangle$. One can compute and simplify the increment as follows,
\bea
\triangle_{p_j} ( \beta_{j+1},\gamma_{j+1} )   & =& A (\beta_{j+1}) (1-2p_j)  \nonumber \\
&& + 2 B (\beta_{j+1},\gamma_{j+1}) \sqrt{p_j - p_{j}^2}   ~~~~ \label{eq:sim} 
\eea
where
\bea
&& A (\beta_{j+1}) = \sin^{2}\frac{\beta_{j+1}}{2}\sin^{2}\theta_{0}, \label{eq:A} \\
&& B(\beta_{j+1},\gamma_{j+1} ) =  \sin \frac{\beta_{j+1}}{2}  \sin\theta_0 \times \nonumber   \\
&& \big( \cos\frac{\beta_{j+1} }{2} \sin \varphi_{j+1} - \sin\frac{\beta_{j+1} }{2}\cos\varphi_{j+1} \cos \theta_0 \big),~\mathrm{and}~~ \label{eq:B} \\
&& \varphi_{j+1}  =  \gamma_{j+1} - \phi_{j+1}. \nonumber
\eea
Since $\sin^2 (\theta_0/2)=1/N$, we note that $A(\beta_{j+1}) = O(N^{-1})$ and $B(\beta_{j+1},\gamma_{j+1}) = O(N^{-1/2})$.

\subsection{  Example: the Grover iteration}

Let us revisit the Grover iteration and investigate its decomposition in Eq. (\ref{eq:sim}). The Grover iteration is the case with $(\beta_j,\gamma_j) = (\pi,\pi)$ for all $j$, by which
\bea
A(\pi) & = & \frac{4(N-1)}{N^2}~\mathrm{and}~ B(\pi,\pi) =  \frac{2 (N-2)\sqrt{N-1}}{N^2}.  \nonumber
\eea
The initial state in Eq. (\ref{eq:initial}) is transformed by $j$ iterations of the Grover operator $G(\pi,\pi)$ \cite{jozsa} to the following state,
\bea
|s_j\rangle = \sin\frac{\theta_0(2j+1)}{2} |t\rangle + \cos \frac{\theta_0(2j+1)}{2} |t^{\perp}\rangle,\nonumber
\eea
from which the target probability is given by,
\bea
p_j = |\langle t | G^j ( \pi, \pi) | s_0 \rangle|^2 = \sin^2 \frac{\theta_0(2j+1)}{2} . \label{eq:probg}
\eea
In Fig. \ref{grover}, the state $|s_j\rangle = G^j(\pi,\pi) | s_0\rangle$ is shown on the path from an initial state to a target one.

Two remarks are in order. Firstly, an increment by a Grover iteration leads to a quadratic speedup depending on a given target probability. For instance, the first increment by the Grover iteration is the following,
\bea
\Delta_{p_0}(\pi,\pi) = \sin^2 \frac{3 \theta_0}{2} -\sin^2 \frac{\theta_0}{2} = \frac{8 (N-1) (N-2)}{N^3}, \nonumber
\eea
which is $O(N^{-1})$. Let us consider a Grover iteration for a target probability around $1/2$, i.e., $p_i=1/2+\epsilon$ for some $\epsilon>0$. It holds that,
\bea
\Delta_{p_i}(\pi,\pi) &=& -2 \epsilon  A(\pi) +  \sqrt{1-4\epsilon^2} B(\pi,\pi) \nonumber 
\eea
which is $O({N^{-1/2}})$. When a target probability close to $1$, e.g., $p_k = 1-\epsilon^{'}$ for $\epsilon^{'}>0$, the increment is given by
\bea
\Delta_{p_k}(\pi,\pi) &=& A(\pi) (2\epsilon^{'} -1) +  2 \sqrt{ \epsilon^{'} (1-\epsilon^{'}) } B(\pi,\pi), \nonumber 
\eea
which in fact highly depends on how small $\epsilon^{'}$ is. For sufficiently small $\epsilon^{'}>0$, the increment becomes negative: this means that the iterations should have stopped. In Fig. \ref{grover}, the increments by Grover iterations are plotted for the case of $8$ qubits.

Secondly, it is crucial that $B(\pi,\pi) = \Omega(N^{-1/2})$ in order to have a quadratic speedup. The increment by a Grover iteration is found as 
\bea
\triangle_{p_j} (\pi, \pi) & = &  \sin^2 \frac{\theta_0(2j+3) }{2} -  \sin^2 \frac{\theta_0(2j+1)}{2}  \nonumber \\
&=& A(\pi) (1-2p_j) + 2 B(\pi,\pi) \sqrt{p_j - p_{j}^2}. \nonumber
\eea 
From Eq. (\ref{eq:probj}), a target probability after $K$ iterations can be found as follows, 
\bea
p_K - p_0 & = & \sum_{j=0}^{K-1}\triangle_{p_j} (\beta_{j+1}, \gamma_{j+1}) \nonumber \\
& = & A(\pi) \sum_{j=0}^{K-1} (1-2p_j) + 2 B(\pi,\pi) \sum_{j=0}^{K-1} \sqrt{p_j - p_{j}^2} \nonumber
\eea
in which one can compute the followings: 
\bea
\sum_{j=0}^{K-1} (1-2p_j) & = & \sum_{j=0}^{K-1} \Big(1 - 2\sin^2\frac{\theta_0(2j+1)}{2} \Big) \nonumber \\
& = & \sum_{j=0}^{K-1} \cos(\theta_0(2j+1)) = \frac{\sin 2K\theta_0}{2\sin\theta_0}, ~~\mathrm{and} \nonumber\\
\sum_{j=0}^{K-1}\sqrt{p_j-p^{2}_i} & = & \sum_{j=0}^{K-1}\sqrt{\sin^2\frac{\theta_0(2j+1)}{2}\cos^{2}\frac{\theta_0(2j+1)}{2} } \nonumber \\
& = & \frac{1}{4}\sum_{j=0}^{K-1} \sin(\theta_0(2j+1))= \frac{\sin^2 K\theta_0}{4\sin \theta_0}.  \nonumber 
\eea
The target probability is thus simplified as follows,
\bea
p_K - p_0 =  A(\pi)\frac{\sin2K\theta_0}{2\sin\theta_0} + B(\pi,\pi)\frac{\sin^2K\theta_0}{2\sin\theta_0}. \label{eq:probtot}
\eea
Let us now suppose that a target probability $p_K$ in the left-hand-side is close to $1$, i.e., for some small $\epsilon>0$,
\bea
p_K = \sin^2\frac{\theta_0(2K+1)}{2} = 1-\epsilon. \label{eq:xx}
\eea
This also means that $\sin2K\theta_0$ is close to $0$. Moreover, since $A(\pi) = O(N^{-1})$ and $B(\pi,\pi) = O(N^{-1/2})$, one can find that Eq. (\ref{eq:probtot}) is dominated by the second term. Then, the numerator is expanded and found that
\bea
\sin^2 K\theta_0 = \Omega( K^2 N^{-1}) \nonumber 
\eea
which holds true since $\sin\theta_0 = O(\sqrt{N})$, i.e., $\theta_0 = \Omega(\sqrt{N})$. Hence, it is shown that
\bea
\frac{\sin^2 K\theta_0}{2\sin\theta_0} = \Omega(K^2 N^{-1/2}). \label{eq:x}
\eea
Thus, in Eq. (\ref{eq:probtot}) the fact that $B(\pi,\pi) = \Omega(N^{-1/2})$ leads to a square-root speedup, i.e., $K=O(\sqrt{N})$.\\

\subsection{  Characterization of QAAOs}
\label{subsec:qaao}

From the previous subsection, it is observed that a quadratic speedup in quantum amplitude amplification is possible since $B(\pi,\pi) = \Omega(N^{-1/2})$. This may be generalized as follows. \\

{\bf Definition}. { \it For an $n$-qubit state having a target probability $p$, a quantum iteration $G(\beta,\gamma)$ in Eq. (\ref{eq:io}) is a QAAO for the state when its increment is positive $\Delta_p (\beta,\gamma)>0 $ and $B(\beta,\gamma) = \Omega(N^{-1/2})$ where $N=2^n$, see Eqs. (\ref{eq:inc}) and (\ref{eq:B}). }\\ 

QAAOs are defined such that their sequential applications can construct a quantum amplitude amplification algorithm with a quadratic speedup. That is, it is aimed to find a sequence of iterations such that with $K=O(\sqrt{N})$,
\bea
|t\rangle =_{\epsilon} G(\beta_K, \gamma_K) G(\beta_{K-1}, \gamma_{K-1}) \cdots G(\beta_1, \gamma_1)  |s_0\rangle ~~\label{eq:id}
\eea
where $=_{\epsilon}$ denotes an $\epsilon$-close distance, i.e., $|a\rangle =_{\epsilon} |b\rangle$ means that $|\langle a| b\rangle|^2 =1-\epsilon$. For instance, the Grover algorithm approximate a target state with $\epsilon = O(N^{-1})$. An exact algorithm appears when $\epsilon=0$ \cite{PhysRevA.64.022307, PhysRevA.62.052304}.



We now show that QAAOs are building blocks to construct quantum amplitude amplification algorithms with a quadratic speedup. Suppose that quantum iterations in Eq. (\ref{eq:id}) are QAAOs. In particular, let $B(\beta,\gamma):= \min_j B(\beta_j,\gamma_j) = \Omega(N^{-1/2})$. It follows that 
\bea
p_K - p_0 &= & \sum_{j=0}^{K-1} \triangle_{p_j} (\beta_{j+1},\gamma_{j+1}) = \Omega(K B(\beta,\gamma)). \label{eq:rel}
\eea
If the target probability $p_K$ in the left-hand-side is close to $1$, it holds that $K = O(\sqrt{N})$. In the following, we also show that a QAAO can be found in a wide range of $(\beta,\gamma)$. \\

\begin{figure*}[t]
	\begin{center}
		\includegraphics[angle=0, width=.95\textwidth]{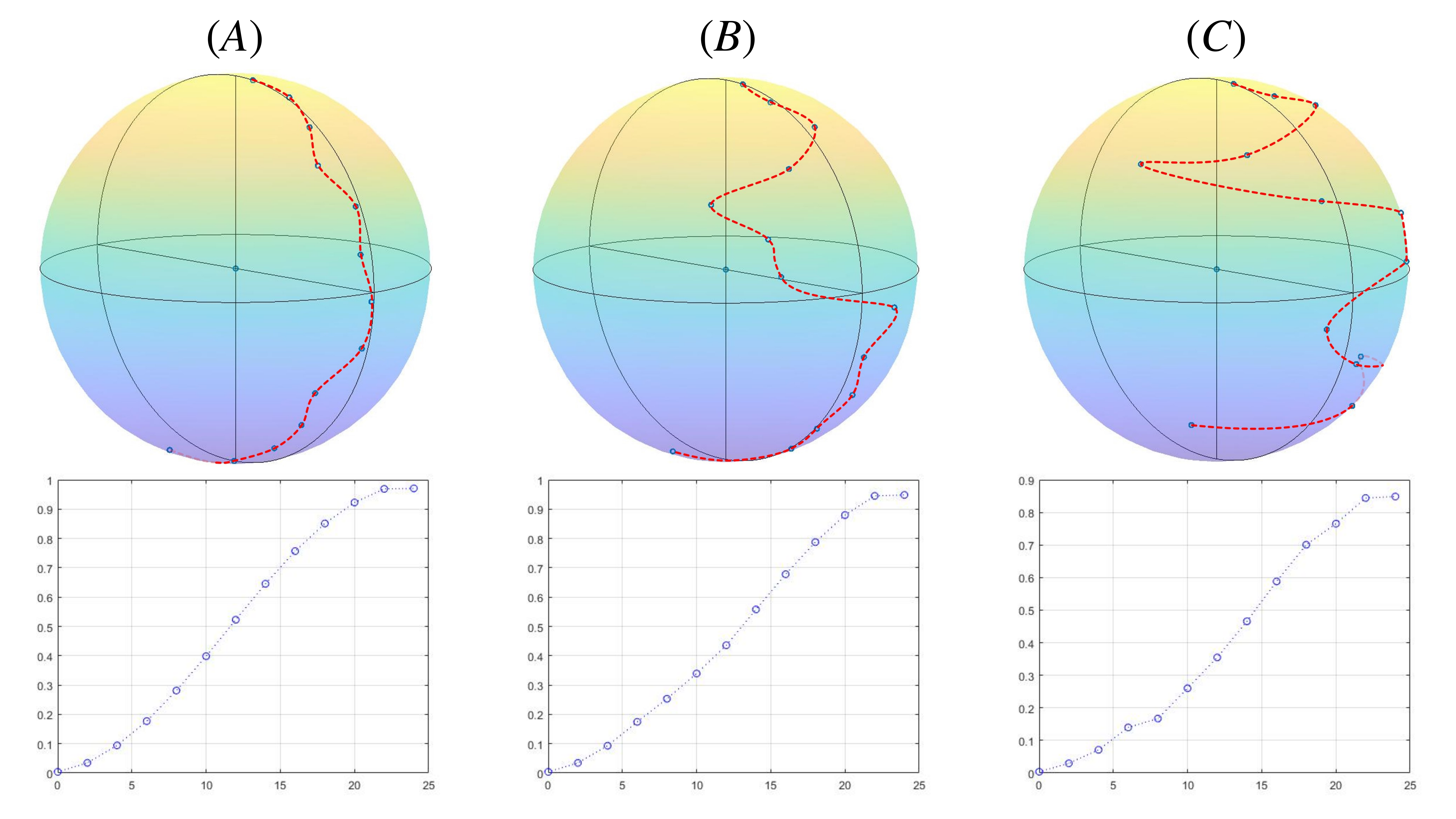}
		\caption{ A sequence of near-optimal QAAOs with parameters randomly chosen from Eq. (\ref{eq:ex}) is considered in cases of $8$ qubits. The $y$-axis is the probability of finding a target state and the $x$-axis the number of oracle queries. Note that the optimal number of the oracle queries is $18$, and here it is about $25$ with the near-optimal QAAOs. (A) A sequence of parameters are chosen by allowing an error $\delta=0.05\pi$. The path on the sphere is not much deviated from the Grover algorithm, see Fig. \ref{grover}. (B) An error is allowed up to $\delta=0.2\pi$, and (C) up to $\delta=0.3\pi$. In all cases, the probability of finding a target keeps increasing and sufficiently high at the end.} \label{qaaof}
	\end{center}	
\end{figure*}

{\bf Proposition 1}. {\it For a sufficiently large $N$, it is probability $1/2$ that randomly chosen parameters $\beta,\gamma\in[-\pi,\pi]$ define a QAAO.} \\

{\it Proof.} Consider a function $f(\beta,\gamma) =1$ if $G(\beta,\gamma)$ is a QAAO and $f(\beta,\gamma) =0$ otherwise. For randomly chosen parameters $\beta,\gamma \in [-\pi,\pi]$, the probability that a QAAO is found is given by,
\bea
\mathrm{Prob}(\beta,\gamma) = \frac{\int_{-\pi}^{\pi}\mathrm{d}\beta\int_{-\pi}^{\pi}\mathrm{d}\gamma ~~ f(\beta,\gamma)}{\int_{-\pi}^{\pi}\mathrm{d}\beta\int_{-\pi}^{\pi}\mathrm{d}\gamma}. \label{Eq:pr}
\eea
The denominator is $4\pi^2$. To compute the numerator, one has to find the range of $(\beta,\gamma)$ where $B(\beta,\gamma)>cN^{-1/2}$ for some $c>0$. Since $N$ tends to be large, we are interested in the converging range. From Eq. (\ref{eq:B}), the condition $B(\beta,\gamma)>0$ is equivalent to the following,
\bea
  \sin\frac{\beta}{2}\sin(\gamma-\phi-\frac{\beta}{2}) >0. \nonumber
\eea
When $\beta\in(0,\pi)$, we have $B(\beta,\gamma)>0$ for $(\gamma-\phi-\beta/2)\in(0,\pi)$. For $\beta\in(-\pi,0)$, it holds $B(\beta,\gamma)>0$ if $(\gamma-\phi-\beta/2)\in(-\pi,0)$.  Then, the numerator of the Eq. (\ref{Eq:pr}) can be computed as
\bea
\int_{0}^{\pi}\mathrm{d}\beta\int_{\phi+\frac{\beta}{2}}^{\phi+\frac{\beta}{2}+\pi}\mathrm{d}\gamma + \int_{-\pi}^{0}\mathrm{d}\beta\int_{\phi+\frac{\beta}{2}-\pi}^{\phi+\frac{\beta}{2}}\mathrm{d}\gamma = 2\pi^2.\nonumber
\eea
Hence, it is shown that $\mathrm{Prob}(\beta,\gamma) = 1/2$. $\Box$\\

\begin{figure*}[t]
	\begin{center}
		\includegraphics[angle=0, width=.95\textwidth]{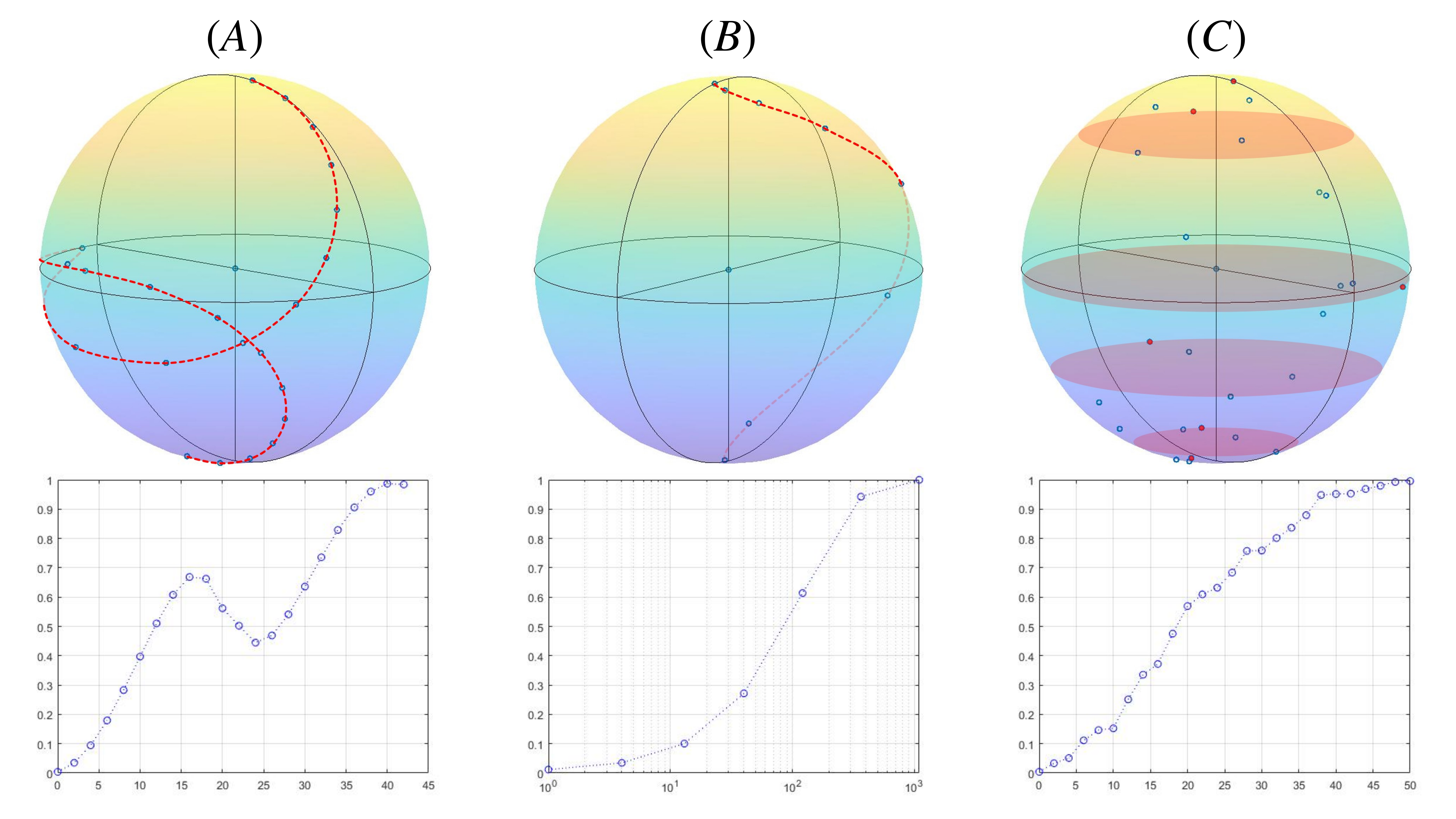}
		\caption{  Quantum amplitude amplification is performed in cases of $8$ qubits. The $x$-axis is the number of oracle uses, and the $y$-axis the probability of finding a target state.  (A) The $\pi/3$-algorithm is plotted \cite{PhysRevLett.95.150501}. The amplitude increases all the time till $10^3$ oracle calls, without a quantum speedup. (B) Fixed-point quantum search with an optimal query complexity is plotted \cite{PhysRevLett.113.210501}. The amplitude of a target state decreases meanwhile, and the oracle is called $45$ times. (C) QAAOs are randomly generated and concatenated so that the amplitude keeps increasing until it reaches to $1$ after the oracle calls $50$ times.} \label{figcomp}
	\end{center}	
\end{figure*}

\begin{figure*}[t]
	\begin{center}
		\includegraphics[angle=0, width=.88\textwidth]{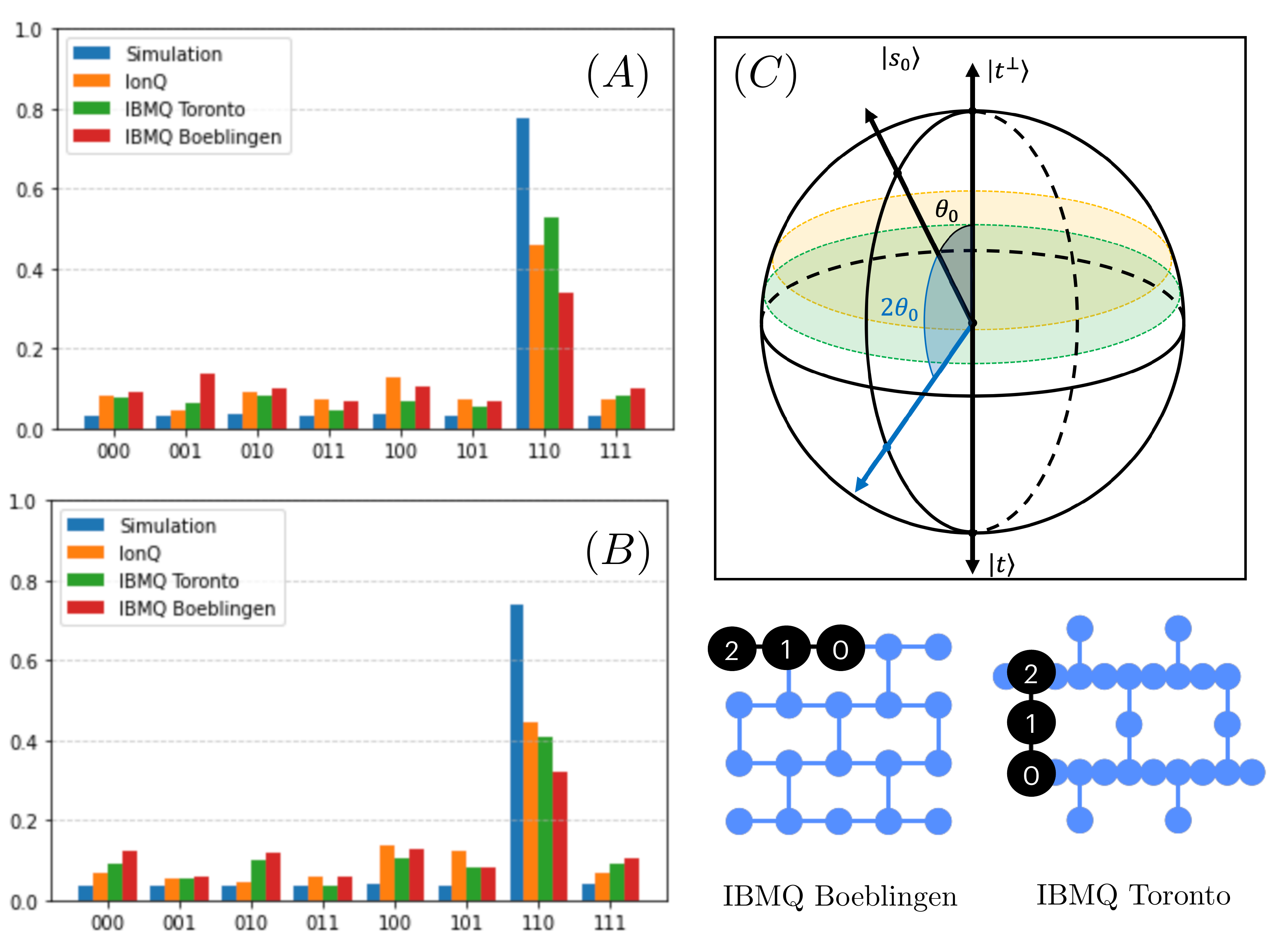}
		\caption{ QAAOs for three-qubit states are realized through cloud-based quantum computing services in IBMQ Boeblingen, IBMQ Toronto, and IonQ. The initial state is prepared as a uniform superposition of $8$ states, and the target state is fixed by $|t\rangle = |110\rangle$. The geometry of qubits in the devices is also shown, where the three qubits lated by $0$, $1$, and $2$ are allocated. (A) The Grover iteration $G (\pi,\pi)$ is performed. (B) A near-optimal QAAO $G(2,634, \pi)$ is performed. Both cases show that the amplitude of the target state is successfully amplified after an iteration. (C) In the sphere of the target state and its complement, amplitude amplification in the ideal case corresponds to a rotation by an angle $2\theta_0$, see the probability of the ideal case in (A). 
			} \label{nisq2}
	\end{center}	
\end{figure*}

\subsection{ Optimal quantum search}

We have so far identified QAAOs that can be generally used to construct quantum amplitude amplification with a quadratic speedup. Among the quantum amplitude amplification algorithms, we derive an optimal one in what follows. In fact, the optimal and exact algorithm corresponds to the Grover iterations followed by a single QAAO additionally.\\


{\bf Proposition 2}. {\it  A maximal increment by a QAAO for a state $|s(\theta,\phi)\rangle$ depends on the range of $\theta$. For $\theta\in [0,\pi- 2 \theta_0)$, the optimal parameters are $\beta=\pm\pi$ and $\gamma=\phi+\pi$. For an initial state $\phi=0$, the condition corresponds to the Grover iteration. For $\theta\in [\pi-2 \theta_0, \pi]$, the optimal parameters are given by
\bea
\beta^* & = & \pm 2\arcsin (\cos\frac{\theta}{2} \csc \theta_0)~~\mathrm{and}~~\nonumber\\
\gamma^* &=& \phi +\pi - \arctan (\cot \frac{\beta^*}{2} \sec\theta_0). \label{eq:opt}
\eea 
For an initial state prepared in a uniform superposition of all states i.e., $\theta_0 = 2\sin^{-1} \sqrt{1/N}$ and $\phi=0$, optimal and exact quantum search is realized by $K^*+1$ optimal QAAOs: 
\bea
|t\rangle = G(\beta^*,\gamma^*)G^{K^*}(\pi,\pi)  |s_0\rangle  \label{eq:optal}
\eea
where $K^* = \lfloor \pi\sqrt{N}/4  -1/2\rfloor$. }    \\

\begin{figure*}[t]
	\begin{center}
		\includegraphics[angle=0, width=.8\textwidth]{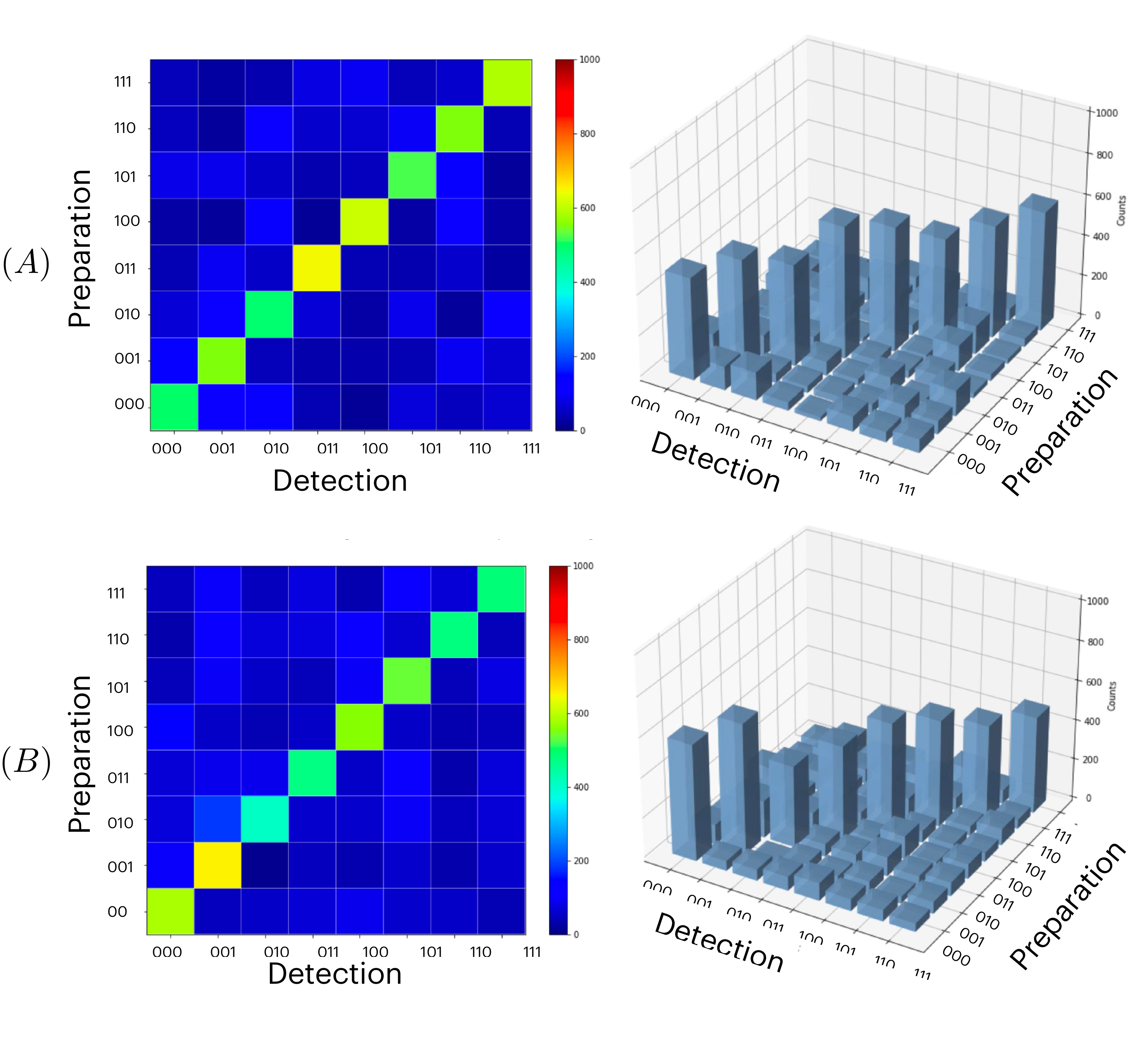}
		\caption{ QAAOs for three-qubit states are realized in the IonQ system. All states $|000\rangle,\cdots,|111\rangle$ are considered as target ones. For instance, when the preparation shows a state $|abc\rangle$ fixed as a target one, the detection shows a resulting state $|a^{'} b^{'} c^{'} \rangle$ for which the amplitude is the most amplified. In (A), a QAAO $G(3.00,-2.98)$ is performed. In (B), a QAAO $G(2.90,3.28)$ is implemented. Both show fairly comparable amounts of amplification of the amplitude of a target state. } \label{ionq}
	\end{center}	
\end{figure*}

{\it Proof}. Let us begin with an arbitrary state $|s(\theta, \phi)\rangle$ for $\theta\in[0,\pi]$ and $\phi\in[0,2\pi]$. Note that the probability of finding a target from the the state is given by $\sin^2(\theta/2)$, meaning that in QAAOs it must be that $\theta\in[0,\pi]$. An increment introduced by $G(\beta,\gamma)$ with $\beta,\gamma\in[-\pi,\pi]$ is denoted by $\Delta (\beta,\gamma)$. For convenience, let us also introduce a parameter  $\varphi=\gamma-\phi\in(0,2\pi)$.  

It is now aimed to find parameters $(\beta,\gamma)$ to define an optimal iteration $G(\beta,\gamma)$. First of all, one can find an optimal $\varphi^{*}$ as follows,
\bea
&& \frac{\partial}{\partial \varphi} \Delta(\beta,\gamma) =0 ~ \iff ~  \tan\varphi^{*} = -\cot\frac{\beta}{2}\sec\theta_{0} \label{eq:con_varphi}.
\eea
With the optimal one $\varphi^{*}$ above, one can also check that 
\bea
\frac{ \partial^2 }{ \partial \varphi^2 } \Delta(\beta,\gamma) \big|_{\varphi^{*}}<0.\nonumber
\eea
This shows the increment is maximal with the parameter $ \varphi^{*} $ in Eq. (\ref{eq:con_varphi})

We recall that for a quadratic speedup, it is necessary to have 
\bea
&&B(\beta, \gamma) > 0,~\mathrm{or ~equivalently } \label{eq:con1} \\
&&\cot\frac{\beta}{2}\sin\varphi - \cos\varphi\cos\theta_{0} > \sin\theta_{0}\cot\theta \label{eq:con2}
\eea
From Eq. (\ref{eq:con_varphi}), the condition above can be simplified as
\bea
&& \varphi^{*}\in\Big(\frac{\pi}{2}, \frac{3\pi}{2}\Big), \label{eq:res_con1}~\mathrm{and} \\
&& \Big|\sin\frac{\beta}{2}\Big| \le \Big|\frac{\sin\theta}{\sin\theta_{0}}\Big|, \label{eq:res_con2}
\eea
respectively. From these, an optimal parameter $\gamma^{*}$ is obtained as follows,
\bea
\gamma^{*}=\phi+\pi-\arctan\Big( \cot\frac{\beta}{2}\sec\theta_{0} \Big). \label{eq:opt_gamma}
\eea
Note that an optimal $\gamma^{*}$ can be defined when a state $|s( \theta,\phi)\rangle$ is identified.

It is now left to find an optimal parameter $\beta^{*}$. Let us introduce $t\in[0, 2\theta_{0}]$ such that
\bea
\sin\frac{t}{2} = \Big| \sin\frac{\beta}{2} \Big| \sin\theta_{0}. \label{eq:t}
\eea
Then, let us rewrite the increment in terms of the optimal parameter $\gamma^{*}$:
\bea
\triangle(\beta,\gamma^{*})=\sin\Big(t+\frac{\theta}{2}\Big)\sin t, \label{eq:inc_opt_gamma}
\eea
which we maximize over $\beta$, see Eq. (\ref{eq:t}).

Similarly to the method we compute $\gamma^*$, one can find an optimal $t^{*}$ as follows, 
\bea
\frac{\partial}{\partial t} \Delta(\beta,\gamma^{*})=0 ~ \iff ~ t^{*}=\pi - \theta \label{eq:con_t}.
\eea
With the optimal one $t^{*}$ above, one can also check that
\bea
\frac{ \partial^2 }{ \partial t^{2} } \Delta(\beta,\gamma^{*}) \big|_{t^{*}}<0\nonumber
\eea
which shows the increment is maximal with the parameter $t^{*}$ in Eq. (\ref{eq:con_t}). From Eq. (\ref{eq:t}) an optimal parameter $\beta^{*}$ can be therefore written as follows,
\bea
\beta^{*} = \pm 2\arcsin\Big( \cos\frac{\theta}{2}\csc\theta_{0} \Big)
\eea
for $\theta\in[\pi-2\theta_{0}, \pi]$. For $\theta\in[0, \pi-2\theta_{0}]$, Eq. (\ref{eq:inc_opt_gamma}) can be rewritten as,
\bea
\triangle(\beta, \gamma^{*})=\frac{1}{2}\Big(\cos\theta-\cos(\theta+t)\Big).\label{eq:conv_inc}
\eea
The increment above is non-negative for all $\theta\in[0,\pi-2\theta_{0})$ and maximized at $t=2\theta_{0}$. From these, an optimal parameter is obtained as $\beta^{*}=\pm\pi$. $\Box$\\

Proposition $2$ in fact reproduces a proof of the optimality of the Grover iteration for those states far from a target one. When a state is closer to a target, i.e., a state $|s(\theta,\phi)\rangle$ with $\theta \geq \pi- 2 \theta_0$ that may be obtained after Grover iterations $K^*$ times, the Grover iteration is no longer a QAAO for the state; by the iteration the probability of finding a target state would decrease. This also explains why the Grover algorithm has to stop right after $K^*$ iterations, at which a measurement therefore reads a target state with an error $O(N^{-1})$. 

For exact and optimal search shown in Eq. (\ref{eq:optal}), an extra iteration $G(\beta^*,\gamma^*)$ is additionally needed. This in fact performs an exact transformation from the state resulting by the Grover algorithm to a target one precisely. We also remark that to achieve an exact transformation, it is essential to exploit an oracle query $e^{-i \gamma |t\rangle \langle t|}$ with $\gamma\neq \pi$, whereas the Grover iteration is with $\gamma=\pi$ all the time.

Let us now consider a set of parameters which are $\delta$-close to optimal ones. As QAAOs can be defined in a wide range of parameters from Proposition $1$, optimal QAAOs may be robust to noise in the preparation of the optimal parameters. These parameters are generated as $(\beta_k,\gamma_k)$ for $1\leq k\leq K^*+1$ by allowing errors up to $\delta$:
\bea
&& \mathrm{for}~~k\in[1,K^* ],~  \beta_k,\gamma_k \in[\pi-\delta, \pi+\delta]\nonumber\\
&& \mathrm{and} ~~| \beta_{K^* +1} -\beta^*| \leq \delta ~ \mathrm{and}~ | \gamma_{K^* +1} -\gamma^*| \leq \delta. \label{eq:ex}
\eea
For instance, cases with $\delta = 0.05\pi, 0.2\pi, 0.3\pi$ are considered and those QAAOs are concatenated. In Fig. \ref{qaaof}, the probability of finding a target state is plotted in the cases of $8$ qubits. The sequences achieve a sufficiently high probability of finding a target state at the end. It is also shown that QAAOs are generically resilient to errors in the preparation of optimal parameters.

\subsection{ Exact quantum search algorithms }

Having identified QAAOs and their optimal sequences for an exact quantum search, we are now in a position to present a generic and systematic scheme of constructing a sequence of QAAOs such that an exact search is achieved with a quadratic speedup. Since a QAAO is characterized by a pair of parameters $(\beta,\gamma)$, we first devise an algorithm that generates a sequence of the parameters 
\bea
S= \{(\beta_1,\gamma_1),~ (\beta_1,\gamma_1),~\cdots,~(\beta_L,\gamma_L)  \} \nonumber
\eea
which ensures an exact search. A generated sequence also finds the number of iterations $L$ containing a quadratic speedup over a classical case. 

We can then compose an exact quantum search algorithm by sequentially applying QAAOs characterized by parameters given the number of iterations too. Note that all exact quantum algorithms can be reproduced as a sequence of QAAOs. In the following, we provide two algorithms, {\bf Parameter Generator} denoted by $P_G$ and {\bf Exact Quantum Search} denoted by $Q$. 


\begin{algorithm}[H]
\begin{algorithmic}[]
\Input{an initial state $|s_{0}\rangle := | s (\theta_0,\phi_0) \rangle  $, constant $c>0$ }
\Output{ parameters $S = \{ (\beta_{i}, \gamma_{i}) \}$}
\Procedure{}{} {$P_G$}{$(|s_{0}\rangle)$}
	\State $i = 0$
	\State $S \gets \emptyset$
	\While {$p_i \neq 1 $}
		\State $i \gets i+1$
		\If {$\theta_i \in [\pi-2\theta_0,\pi]$}
			\State $\beta_i \gets \beta^*$
			\State $\gamma_i \gets \gamma^*$
		\Else
			\Do
			\State generate $(\beta_{i}, \gamma_{i})$ for a state $|s(\theta_{i-1}, \phi_{i-1})\rangle$
			\doWhile{$B(\beta_{i}, \gamma_{i}) \leq cN^{-1/2}$ \textbf{or} $\Delta_{p_{i-1}}\le 0$}
		\EndIf
		\State $|s(\theta_i,\phi_i)\rangle \gets G(\beta_i,\gamma_i)|s(\theta_{i-1},\phi_{i-1})\rangle$
		\State $S \gets S \cup \{(\beta_i,\gamma_i)\}$
	\EndWhile
\State \Return $S$
\EndProcedure
\end{algorithmic}
\caption{ {\bf Parameter Generator} ($P_G$)}
\end{algorithm}

Then, with a set of parameters obtained from {\bf Parameter Generator} $P_G$, an exact quantum search algorithm can be constructed as follows. Note that the cardinality of the set $S$ obtained in the algorithm $P_G$ defines the number of iterations, denoted $L$, for exact quantum search. 

\begin{algorithm}[H]
\begin{algorithmic}[]
\Input{an initial state $|s_{0}\rangle := | s (\theta_0,\phi_0) \rangle  $ }
\Output{a target state $|t\rangle$}
\Procedure{}{} {$Q$}{$(|s_{0}\rangle)$}
	\State $S \gets P_G(|s_{0}\rangle)$
	\State $L \gets |S|$
	\State \Return $G(\beta_{L },\gamma_{L }) \cdots G(\beta_{1 },\gamma_{1 })  |s_0\rangle$
\EndProcedure
\end{algorithmic}
\caption{{\bf Exact Quantum Search} (Q) }
\end{algorithm}

In Figs. \ref{qaaof} and \ref{figcomp} (C), we generate QAAOs for $8$-qubit states and plot the evolution of an initial state to a target one. It is shown that the amplitude of a target state keeps increasing all the way through. 

\subsection{ Comparison to fixed-point quantum search}

Having identified a quantum iteration for amplitude amplification, we revisit fixed-point quantum search algorithms. In the $\pi/3$-algorithm, an initial state converges arbitrarily to a target one by the iterations. Thus, it is not necessary to know when an algorithm has to stop. However, it takes $O(N)$ times that a target state is found with a sufficiently high probability \cite{PhysRevLett.95.150501}. Then, it is clear that some of the iterations are not QAAOs. In Fig. \ref{figcomp} (A), the probability of finding a target state is plotted and it is shown that the probability keeps increasing by the iterations, some of which do not correspond to a QAAO.


It is remarkable that a fixed-point search algorithm with a quadratic speedup has been presented \cite{PhysRevLett.113.210501}. The algorithm is structured as it is shown in Eq. (\ref{eq:io}), and starts by fixing a lower bound to the probability of obtaining a target state at the end. From the bound, a minimal number of iterations $L = O(\sqrt{N})$ are provided as well as the sequence of parameters $(\beta_j,\gamma_j)$ specifically via the Chebyshev polynomials, see Appendix for the detail.

Then, a measurement readout after $L^{'}$ iterations with $L^{'}>L$ finds a target state with a probability higher than the bound fixed in the beginning. Note that, after $L$ iterations, the probability fluctuates within the bound. However, as soon as exact search is attempted, i.e., by putting the lower bound to $1$, a quadratic speedup disappears. 

In Fig. \ref{figcomp} (B), the probability of finding a target state is plotted for $8$ qubits. It is shown that the probability decreases meanwhile during some of the iterations, see Appendix for the details and also Table \ref{tablem}. This shows that the algorithm is not a sequence of QAAOs. Thus, we have shown that none of the fixed-point algorithms so far can be realized by QAAOs only. One may assert that it is necessary to include iterations that are not QAAOs to compose a fixed-point quantum search algorithm.

\begin{table}[H]
\centering
\label{tab:fpa_result}
\begin{tabular}{@{}cllllllllllc@{}}
\hline
    {\bf No.} & \vline& {\bf $\mathrm{State}$:~($\theta$, $\phi$)} & \vline& {\bf  $\mathrm{Iteration}$: ($\beta$, $\gamma$)} &\vline  && {\bf $\triangle(\beta, \gamma)$}    \\ 
\hline
\hline
    9 &\vline& (1.9147, 5.1123)& \vline& (2.8209, 2.895)& \vline& &-0.0061  \\
    10 &\vline& (1.9018, 4.4555)& \vline& (2.4078, 2.6915)& \vline& &-0.1007    \\
    11 &\vline& (1.6947, 3.2412)& \vline& (-1.4255, 1.4255)&\vline && -0.0596    \\
    12 &\vline& (1.5752, 3.2562)& \vline& (-2.6915, -2.4078)& \vline&& -0.0575   \\
    \hline
\end{tabular}
\caption{ Parameters in the fixed-point quantum search algorithm with an optimal number of oracle queries are shown for cases of $8$ qubits. The increment $\Delta$ is negative from the $9$th to $12$th iterations. This shows the algorithm is not a sequence of QAAOs.  }\label{tablem}
\end{table}

\section{Realization of QAAOs in cloud-based quantum computing}

QAAOs are realized in cloud-based quantum computing services provided by IBMQ and IonQ, in particular, IBMQ Toronto, IBMQ Boeblingen, and IonQ. The IBMQ devices implement superconducting qubits arrayed in a fixed geometry, see Fig. \ref{nisq2}. The IonQ system applies trapped ion qubits, which do not form a fixed geometry. In both systems, circuits of QAAOs are designed as shown in Fig. \ref{circuit}, and single iterations of QAAOs are realized. 


In Fig. \ref{nisq2}, the results with IBMQ performed on Jan. 8 2021 and with IonQ on Jan. 7 2021 are shown. The target state is fixed by $|110\rangle$. All of the systems in IBMQ and IonQ show fairly comparable amount of amplification of the amplitude of the target state. In Fig. \ref{ionq}, the realizations of QAAOs for different target states in IonQ are presented. These are performed on Mar. 16 2021. Two near-optimal QAAOs are performed. 

The demonstrations in IBMQ and IonQ have shown that three-qubit QAAOs can be realized in the cloud-based quantum computing services. In all cases, fairly comparable performance of QAAOs is presented. All these show that a single iteration of QAAO is feasible with currently available NISQ technologies. It may be anticipated that QAAOs can be used to quantum computing applications in practice. 

\section*{Conclusion}

In conclusion, we have characterized QAAOs, the quantum iterations that can be generally used to construct quantum amplitude amplification with a quadratic speedup. Our results show that QAAOs can be identified in a wide range of parameters in such a that randomly chosen parameters can construct a QAAO with probability almost $1/2$. Thus, on the one hand, it is feasible to prepare sequences of QAAOs. On the other hand, QAAOs are generically resilient to errors appearing in the preparation of optimal parameters. 

QAAOs for three qubits are realized through the cloud-based quantum computing services in IBMQ and IonQ. The results show fairly comparable amounts of amplification of target states. Thus, a single iteration of a QAAO with suitable parameters is feasible with the current quantum technologies. We envisage that QAAOs may be applied in practical quantum computing applications. 

We remark, as mentioned above, our results generalize exact quantum algorithms. In fact, all of the exact quantum search algorithms with a quadratic speedup, known so far, can be reproduced as sequences of QAAOs. It turns out that the Grover iterations are optimal when an evolving state is not sufficiently close to a target one. Interestingly, it is shown that none of the existing fixed-quantum algorithms can be a sequence of QAAOs. The $\pi/3$-algorithm has no quantum speedup, from which it is clear that some of the iterations are not QAAOs. The fixed-point quantum search with a quadratic speedup consists of some iterations by which the probability of a target state actually decreases. This shows that the fixed-point algorithm is not a series of QAAOs and also cannot be optimal. Thus, distinctions between exact and fixed-point algorithms may be elucidated in terms of QAAOs.  

In future investigations, on the practical side, it is interesting to optimize quantum circuits of QAAOs, e.g. \cite{PhysRevA.101.032346}, and to make them fitted to quantum computing applications in practice, e.g., optimization \cite{10.1137/040605072, optimizaiton}, state preparation \cite{Soklakov:2005wx}, high-energy physics \cite{PhysRevD.101.094015}, cryptanalysis \cite{8490203, 10.1007/978-3-319-29360-8_3, PMCID:PMC7991734}, etc. From a fundamental point of view, it would be interesting to seek the characterization of quantum iterations for fixed-point quantum search. It is left open to construct a fixed-point quantum search algorithm that contains both of the properties i) finding a target with certainty and ii) having a quadratic speedup. 


\section*{Acknowledgement}
This work is supported by National Research Foundation of Korea (NRF-2019M3E4A1080001), the ITRC (Information Technology Research Center) Program (IITP-2021-2018-0-01402) and Samsung Research Funding $\&$ Incubation Center of Samsung Electronics (Project No. SRFC-TF2003-01).


%
\newpage

\section*{Appendix : Fixed-point quantum search with an optimal number of queries}

The fixed-point quantum search algorithm with an optimal number of queries \cite{PhysRevLett.113.210501} works even if the {\it a priori} information $\lambda = |\langle t | s(\theta_0, \phi_0) \rangle|$ for an initial state $| s(\theta_0, \phi_0) \rangle$ is not provided, i.e., quantum counting that verifies $\lambda$ is not needed. One has to fix a lower bound to the probability of finding the target at the end: let $\delta^2$ denote an error rate in the worst case so that the probability of a target at the end is not lower than $1-\delta^2$. This also tells least number of iterations $K^*$ such that for all $K^{'}\geq K^*$ one can have $P_{K^{'}}\geq 1-\delta^2$. 


Given an error bound $\delta^2$, the parameters $(\beta_i, \gamma_i)$ of $K^*$ iterations are given by Chebyshev polynomials. The $k$-th Chebyshev polynomial of the first kind is denoted by $T_k (x)=\cos\Big( k  \cos^{-1}(x)\Big)$. Then, the parameters $(\beta_i,\gamma_i)$ are given by
\bea
\beta_i=\gamma_{K^* - i + 1}=2\cot^{-1}\Big( \tan(2\pi i / K^*)\sqrt{1-\eta^2} \Big) \label{eq:fpa_pg}
\eea
where $\eta^{-1}=T_{1/K^*}(1/\delta)$. 

\begin{figure}[H]
	\begin{center}
		\includegraphics[angle=0, width=.47\textwidth]{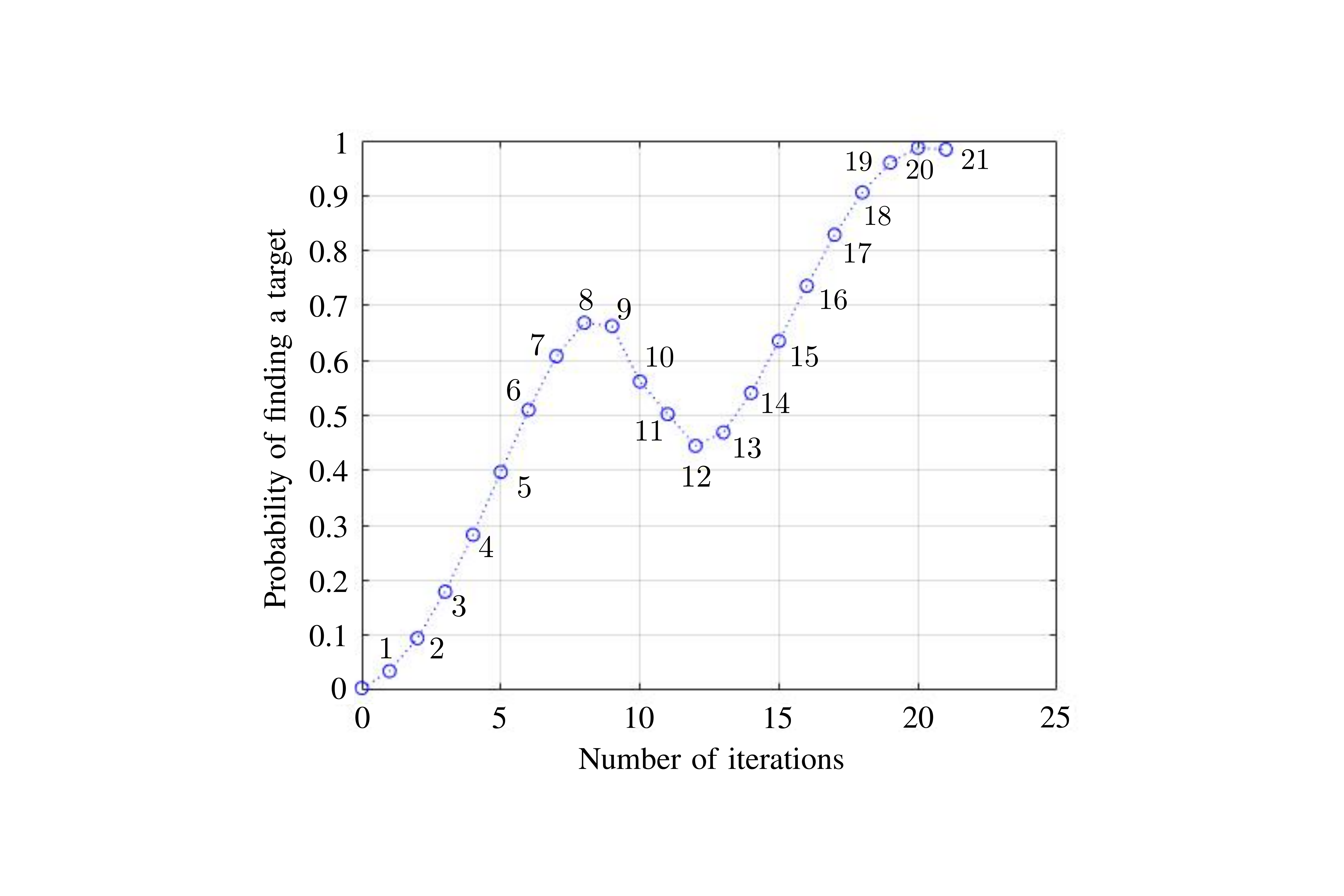}
		\caption{ The probability of finding a target is plotted in the fixed-point algorithm showin in \cite{PhysRevLett.113.210501} for $8$ qubits. The $x$-axis denotes the number of iterations, in which each iteration exploits an oracle query twice. The iterations for the $9$th to the $12$th steps are not QAAOs for the states under evolution. Numerical data are shown in Table \ref{table}. } \label{fixedex}
	\end{center}	
\end{figure}

The algorithm for $8$ qubits and a single target item is considered. In Table \ref{table}, the parameters $(\beta_i,\gamma_i)$ are shown. The error bound is taken as $\delta=0.316$, from which the sequence of parameters are generated from Eq. (\ref{eq:fpa_pg}) and listed in the table. Then, one has $P_{K^{'}} \geq 0.9$ for $K^{'} \geq 20$, see also Fig. \ref{fixedex}. \\


\begin{table}[H]
\centering
\label{tab:fpa_result}
\begin{tabular}{@{}cllllllllllc@{}}
\hline
    {\bf No.} & \vline& {\bf $\mathrm{State}$:~($\theta$, $\phi$)} & \vline& {\bf  $\mathrm{Iteration}$: ($\beta$, $\gamma$)} &\vline  && {\bf $\triangle(\beta, \gamma)$} &\vline& { QAAO}  \\ 
\hline
\hline
    1 &\vline& (0.1251, 0)         & \vline &  (3.1291, 3.1354) &\vline&  &0.0309 &\vline& O  \\
    2 &\vline& (0.3752, 6.2727)&\vline &  (3.1162, 3.1228) &\vline & &0.0598 &\vline& O \\
    3 &\vline& (0.6252, 6.244)  & \vline & (3.102, 3.1093) & \vline& &0.0847 &\vline& O \\
    4 &\vline& (0.8746, 6.1934) &\vline & (3.0857, 3.0941) & \vline& &0.1036 &\vline& O \\
    5 &\vline& (1.1218, 6.1143) &\vline & (3.0659, 3.0763)& \vline& &0.1141 &\vline& O \\
    6 &\vline& (1.3633, 5.9948)& \vline& (3.0401,  3.0539)&\vline & &0.1133 &\vline& O \\
    7 &\vline& (1.5914, 5.8145)& \vline& (3.0033, 3.0235)& \vline& &0.0975 &\vline& O \\
    8 &\vline& (1.7881, 5.5385)&\vline & (2.9433, 2.9775)& \vline& &0.0608 &\vline& O \\
    9 &\vline& (1.9147, 5.1123)& \vline& (2.8209, 2.895)& \vline& &-0.0061 &\vline& X \\
    10 &\vline& (1.9018, 4.4555)& \vline& (2.4078, 2.6915)& \vline& &-0.1007 &\vline& X \\
    11 &\vline& (1.6947, 3.2412)& \vline& (-1.4255, 1.4255)&\vline && -0.0596 &\vline& X \\
    12 &\vline& (1.5752, 3.2562)& \vline& (-2.6915, -2.4078)& \vline&& -0.0575 &\vline& X \\
    13 &\vline& (1.46, 4.4549)& \vline&  (-2.895, -2.8209)& \vline&& 0.0245 &\vline& O \\
    14 &\vline& (1.5091,  5.0453)&\vline & (-2.9775, -2.9433)&\vline & &0.0719 &\vline& O \\
    15 &\vline& (1.653, 5.4069)& \vline& (-3.0235, -3.0033)& \vline& &0.0946 &\vline& O \\
    16 &\vline& (1.8456, 5.6353)& \vline& (-3.0539, -3.0401)& \vline& &0.1001 &\vline& O \\
    17 &\vline& (2.0619, 5.7743)&\vline & (-3.0763, -3.0659)&\vline & &0.0931 &\vline& O \\
    18 &\vline& (2.2888, 5.8423)& \vline& (-3.0941, -3.0857)&\vline && 0.077 &\vline& O \\
    19 &\vline& (2.518, 5.834)& \vline& (-3.1093,  -3.102)&\vline & &0.0541 &\vline& O \\
    20 &\vline& (2.7389, 5.6917)& \vline& (-3.1228, -3.1162)&\vline & &0.0269 &\vline& O \\ 
    21 &\vline& (2.912, 5.1506)& \vline& (-3.1354, -3.1291)&\vline & &-0.0028 &\vline& X \\ 
    \hline
\end{tabular}
\caption{The fixed-point algorithm in \cite{PhysRevLett.113.210501} is performed for $8$ qubits. An initial state is prepared as $H^{\otimes 8} |0\rangle^{\otimes8} $ from which the initial condition is given by, $\theta_0 = 2 \sin^{-1} \sqrt{ 2^{-8} } \approx 0.1251 $ and $\phi_0=0$, see Eq. (\ref{eq:state}). For the state, an iteration with $(\beta_1,\gamma_1) =  (3.1291, 3.1354)$ is applied and the amplitude increases by $\triangle=0.0309$. It is shown that for the increment is negative, i.e., $\Delta<0$, from the $9$th to $12$th iterations.  }\label{table}
\end{table}

\end{document}